\documentclass{article}
\title{Transport in randomly-fluctuating spatially-periodic potentials }
\author{ James P. Gleeson\\
 \emph{Department of Mathematics \& Statistics, University of Limerick, Ireland.}}
\date{9 May, 2008}
\usepackage{amsmath}
\usepackage{amsbsy}
\usepackage{amscd}
\usepackage{amsopn}
\usepackage{amstext}
\usepackage{amsxtra}
\usepackage{epsfig}
\def\lesssim{\mathrel{\hbox{\rlap{\hbox{\lower4pt\hbox{$\sim$}}}\hbox{$<$}}}}

\def\be{\begin{equation}}

\def\s{\sigma}

\begin{document}
\maketitle

\begin{abstract}
The motion of overdamped particles in a one-dimensional
spatially-periodic potential is considered. The potential is also
randomly-fluctuating in time, due to multiplicative colored noise
terms, and has a deterministic tilt. Numerical aimulations show two
distinct parameter regimes, corresponding to free-running
near-deterministic particles, and particles which are trapped in
local minima of the potential with intermittent escape flights.
Perturbation and asymptotic methods are developed to understand the
drift velocity and diffusion coefficient in each parameter regime.
\end{abstract}
\section{Introduction} \label{sect1}
The subject of this paper is the statistical characterization of the
motion of particles whose (one-dimensional) position $x(t)$ is
governed by the Langevin equation
\begin{equation}
\frac{d x}{d t} = U + f(t) \cos(k x) + g(t) \sin (k x) \label{en},
\end{equation}
where $f(t)$ and $g(t)$ are random functions of time. The positive
constant $U$ is the deterministic particle velocity, i.e., the speed
at which particles would travel in the absence of the stochastic
terms, and $k$ is a spatial frequency which governs how the noise
sources affects the particle at position $x$. This equation may be
viewed as describing overdamped
 one-dimensional motion in a spatially-periodic potential which fluctuates
randomly in time, with an additional tilt. Denoting the potential as
$V(x,t)$, the equation of motion is
\begin{equation}
\dot x=\frac{ d x}{d t} = - \frac{\partial}{\partial x} V(x,t).
\label{e1}
\end{equation}
with the potential given by
\begin{equation}
V(x,t) = - U x - f(t) \frac{1}{k} \sin(k x) +  g(t)\frac{1}{k}
\cos(k x). \label{e2}
\end{equation}
Here $U$ is interpreted as the constant tilt of the potential.

Equations of motion of this form arise in a number of diverse
applications. In microelectronic circuit design applications, for
example, it is important to accurately predict the effect of
external noise sources upon the phase angle $x(t)$ of nonlinear
self-sustaining oscillators \cite{Demir,Demircolored}. Under some
assumptions, the  equation of motion for $x(t)$ has been shown
\cite{Gleeson_IEEE06, ODoherty_IEEE07} to be of the form (\ref{en}),
with the tilt  $U$ representing the designed oscillator frequency in
the absence of noise, and noise sources (voltage or current sources)
being denoted by $f(t)$ and $g(t)$. The noise sources cause a
diffusive drift of the phase of the oscillator, known  as
\emph{phase noise}. Noise of  a finite bandwidth (colored noise) may
also induce a shift in the average oscillator frequency
\cite{ODoherty_IEEE07}. It has recently been demonstrated
\cite{ODoherty_IEEE07} that a stochastic perturbation method may be
applied to accurately predict the effects of the noise on the
oscillator dynamics in the limit where $U$ is much larger than the
noise intensity, this being the limit of interest for engineering
applications.

Further examples of models of the form (\ref{en}) occur in the study
of weakly coupled
 phase oscillators \cite{Kuramoto}, where the noisy terms $f(t)$ and
$g(t)$ describe fluctuations in the mean field of the oscillators
\cite{Gleeson_EPL06}, and in an agent-based model of stock market
price movements \cite{Gleeson_PhysicaA05}. For all these examples,
information on the qualitative behavior of the first and second
moments of the process $x(t)$ described by equation (\ref{en})  is
very desirable. In this paper we therefore examine the mean velocity
and diffusion coefficient for an ensemble of trajectories $x(t)$
 in all regimes of parameter space, using extensive numerical
 simulation, perturbation approaches, and appropriate
 non-perturbative methods.

The random functions $f(t)$ and $g(t)$ appearing in equation
(\ref{e2}) are independent, stationary, zero-mean Gaussian
processes, which are characterized by their variance $\alpha^2$ and
correlation function $R(t)$:
\begin{equation}
\left< f(0)f(t)\right> = \left< g(0)g(t)\right> =\alpha^2 R(t).
\label{e4}
\end{equation}
(Angle brackets are used throughout the paper to denote ensemble
averages). We briefly consider the case where the stochastic terms
$f(t)$ and $g(t)$ are white noise sources (so that $R(t)$ is a delta
function), but most of our analysis is focused on coloured noise,
where the presence of a non-zero correlation timescale $\tau$ for
the noise has several nontrivial effects. Note that we do not add an
independent white noise term to the right hand side of equation
(\ref{e1}), as used to model thermal diffusion in
\cite{Reimann_PRL01} for example. Neglecting such diffusion effects
at this stage allow us to clearly demonstrate the role of the
fluctuations in the potential; moreover, it is expected that low
levels of additive white noise will not qualitatively affect the
results determined here.

\begin{figure}
\centering
 \epsfig{figure=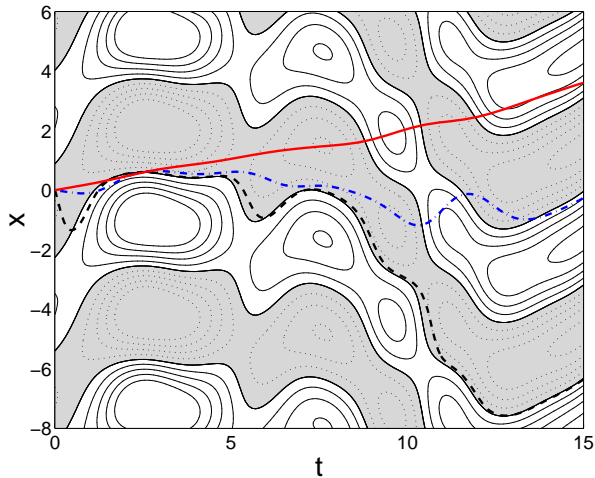,width=8.0cm}
\caption{Sample trajectories from equation (\ref{e8}) shown with a
single realization of the random field $f(t) \cos x + g(t) \sin x$.
The background contours and shading represent the values of the
random field (negative field values are shaded, with dotted
contours), and the lines show the solutions $x(t)$ of equation
(\ref{e8}) with $\mu=0.25$ and $\kappa=0.1$ (red solid line),
$\kappa=1$ (blue dot-dashed line), and $\kappa=10$  (black dashed
line). }\label{figtestcontour}
\end{figure}
\begin{figure}
\centering
 \epsfig{figure=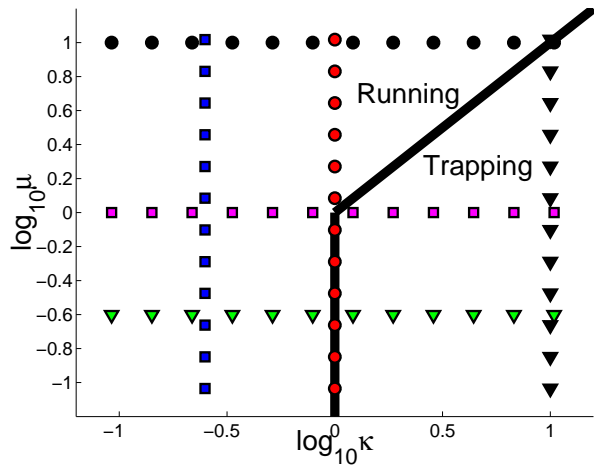,width=8.0cm}
\caption{Schematic of the $\kappa$--$\mu$ parameter plane, showing
the boundary $\kappa=\text{max}(1,\mu)$ between the running regime
(which lies to the left of the solid black line) and the trapping
regime (to the right of the line) as given in equation
(\ref{runcond}). Symbols indicate the lines of constant-$\kappa$ and
constant-$\mu$ along which the simulation results of Figures
\ref{fig1} and \ref{fig1b} are collected.}\label{figparamplane}
\end{figure}

Figure \ref{figtestcontour} provides an illustration of the effects
of various parameters in the problem. These parameters are discussed
in detail in section \ref{sect3} below, but the basic features are
clear from the three $x(t)$ trajectories  plotted in Figure
\ref{figtestcontour}. If the potential tilt is sufficiently large to
dominate the fluctuating part of the potential, then the particle
moves essentially according to $x(t)\approx U t$, with small
corrections due to the influence of the fluctuating potential. The
solid red line illustrates a realization of such a case, which we
class as being in the \emph{running regime}. On the other hand, if
the tilt is relatively weak compared to the fluctuating-in-time
periodic-in-space potential, the motion of the particle is dominated
by the latter. Indeed, for sufficiently slow fluctuations, we expect
the particle to become trapped in a local minimum of the
 potential, with infrequent escapes to other
nearby energy wells. The trajectory $x(t)$ plotted as the black
dashed line illustrates an example of this so-called \emph{trapping
regime}. Note that the contours and shading represent the gradient
of the fluctuating part of the potential (shading and dotted
contours denote negative values, while solid contours and white
background indicate positive values), and the trajectory $x(t)$
closely follows the position of a local minimum (zero gradient
contour). We defer a full explanation until appropriate
nondimensional parameters are introduced in section \ref{sect3}, but
note that our main  result is the theoretical analysis of motion in
the two asymptotic parameter regimes---running and
trapping---introduced here, and depicted schematically on the
parameter plane of Figure \ref{figparamplane}.

The  remainder of this paper is structured as follows. In section
\ref{sect2} we examine the white-noise limit for the random
processes $f(t)$ and $g(t)$ in equation (\ref{e2}). More interesting
results emerge when the noise processes have a non-zero correlation
time $\tau$; the analysis of such colored noise cases begins with
the introduction of nondimensional parameters in section
\ref{sect3}. Methods of numerical simulation are detailed in section
\ref{sect4} and  theoretical analysis of the two main parameter
regimes is undertaken in sections \ref{sect5} and \ref{sect6}.
Section \ref{sect7} concludes with a discussion of results.

\section{White noise}
\label{sect2} If the noise sources $f(t)$ and $g(t)$ in equation
(\ref{en}) are white (i.e., $R(t)=\delta(t)$ in equation
(\ref{e4})), then the probability density function $P(x,t)$ giving
the particle concentration at position $x$ at time $t$  is the
solution of a Fokker-Planck equation \cite{Risken}. Some care is
needed when interpreting this white-noise limit, as the noise
sources appear in equation (\ref{en}) in a multiplicative fashion.
It is typical in such problems that the It\^{o} and Stratonovich
interpretations of the white noise terms in equation (\ref{e2}) lead
to different results. Rather remarkably however, in this particular
example the same Fokker-Planck equation is found using both
interpretations, as the symmetry of the two noise sources causes the
so-called spurious drift term to vanish. Thus either the It\^{o} or
Stratonovich interpretation of the white noise limit of equation
(\ref{en}) leads to the following equation for $P(x,t)$:
\begin{equation}
\frac{\partial P}{\partial t} + U \frac{\partial P}{\partial x} -
\frac{\alpha^2}{2}\frac{\partial^2 P}{\partial x^2}=0. \label{e5}
\end{equation}
It is straightforward to determine from this equation that the mean
value of $x(t)$ grows at a constant rate $U$; indeed, multiplying
equation (\ref{e5}) by $x$ and integrating over space yields the
exact relation
\begin{equation}
\left<x(t)\right> = \left<x(0)\right> + U t,
\end{equation}
so that the mean velocity is just $U$. Similarly, the diffusion
coefficient of $x$ equals the coefficient of the final term of
equation (\ref{e5}), i.e. $\alpha^2/2$. Note this coefficient shows
no dependence on the value of the potential tilt parameter $U$. As
we show below, qualitatively very different results (including
diffusion coefficients which depend on $U$) are found when the noise
sources  have a non-zero correlation timescale. We therefore proceed
to consider the case of colored noise in the next section.

\section{Colored noise and nondimensional parameters}
\label{sect3} A physically more realistic model of random
fluctuations than the infinitely-fast variation of white noise is
provided by colored noise, for which the correlation function $R(t)$
defined in equation (\ref{e4}) decays from its peak value $R(0)=1$
to zero over some characteristic timescale $\tau$. One example of
such a correlation function  is
\begin{equation}
R(t)=e^{-\frac{t^2}{2 \tau^2}}. \label{e7}
\end{equation}
This correlation function is particularly convenient for numerical
simulation, and also has the property $R^\prime(0)=0$ which is
important for some of our theoretical analysis below. This
smoothness of $R$ at zero argument is related to the finiteness of
the variance of the processes $f^\prime(t)$ and $g^\prime(t)$, but
it is not always guaranteed: if $f(t)$ and $g(t)$  are
Ornstein-Uhlenbeck processes, for example, then
  this smoothness condition is violated. Nevertheless,
for the purposes of this initial examination of the properties of
the equation (\ref{e1}) we assume the smoothness property when
needed, and use the correlation function (\ref{e7}) in all numerical
simulations.

Having determined a characteristic timescale $\tau$, it proves
convenient to introduce dimensionless variables (temporarily denoted
by tildes) by measuring time in units of $\tau$ and space in units
of $1/k$. The variance $\alpha^2$ of the noise sources has
dimensions of (length/time)$^2$, while the tilt $U$ has dimensions
length/time. The dimensionless equation of motion for $\tilde
x(\tilde t)$ may therefore be written as
\begin{equation}
\frac{d \tilde x}{d \tilde t} = \tilde \mu + \tilde\kappa
\left[\tilde f(\tilde t) \cos \tilde x  + \tilde g(\tilde t) \sin
\tilde x\right], \label{e8}
\end{equation}
where  the dimensionless tilt is
\begin{equation}
\tilde \mu=U k \tau
\end{equation}
 and the \emph{Kubo number} \cite{VanKampen} is defined as
 \begin{equation}
 \tilde \kappa = \alpha k \tau.
 \end{equation}
 Note the dimensionless noise sources $\tilde f$ and $\tilde g$ in equation (\ref{e8})
  have unit variance and unit (dimensionless) correlation
time. The physical meaning of the Kubo number can be understood best
in the case of zero tilt, i.e., with $U=0$ in  equation (\ref{en}).
Then particles moving in the fluctuating potential  will travel a
(dimensional) distance of order $\alpha \tau$ during the correlation
time of the noise processes; the Kubo number is the ratio of this
lengthscale to the reciprocal spatial frequency $k^{-1}$ of the
potential. Thus the $\tilde\kappa\to 0$ limit, for example,
corresponds to a potential which fluctuates rapidly in time, so that
a particle moving in it travels a very short distance (compared to
the period of the potential) during the time in which values of
$f(t)$ and $g(t)$ change significantly. On the other hand, taking
$\tilde \kappa \to \infty$ corresponds to a frozen (quenched) limit,
where particles feel the spatial variation of the potential much
sooner than the temporal variation.

For the remainder of the paper we concentrate on the dimensionless
form of the equation (\ref{e8}) and henceforth drop the tilde
notation. We focus particularly on the range of behaviors given by
varying the two  positive parameters $\mu$ and $\kappa$. Results
will be presented for the large-time mean velocity $v$ and the
diffusion constant $D$ of an ensemble of particles. The mean
velocity is defined as
\begin{equation}
v=\lim_{t\to\infty} \frac{d }{d t}\left<x(t)\right>, \label{vdefn}
\end{equation}
or equivalently by the large-time asymptotic form of the mean
position:
\begin{equation}
 \left< x(t)\right> \sim v t \quad \text{   as   } \quad
 t\to\infty \label{vasym}.
\end{equation}
Assuming diffusive behavior over sufficiently long timescales, the
diffusion constant is defined by the large-time asymptotic relation
\begin{equation}
\text{var}(x) \sim 2 D t\quad \text{   as   } \quad t\to
\infty,\label{Ddefn}
\end{equation}
where $\text{var}(x)$ is the variance of the distribution of
particle positions at time $t$:
\begin{equation}
\text{var}(x) = \left< x^2(t)\right> - \left<x(t)\right>^2.
\end{equation}
Alternatively, we may use the limit definition
\begin{equation}
D=\frac{1}{2}\lim_{t\to\infty}\frac{d}{dt}\text{var}(x).
\label{Dlim}
\end{equation}
\section{Numerical simulations}
\label{sect4}

\begin{figure}
\centering
 \epsfig{figure=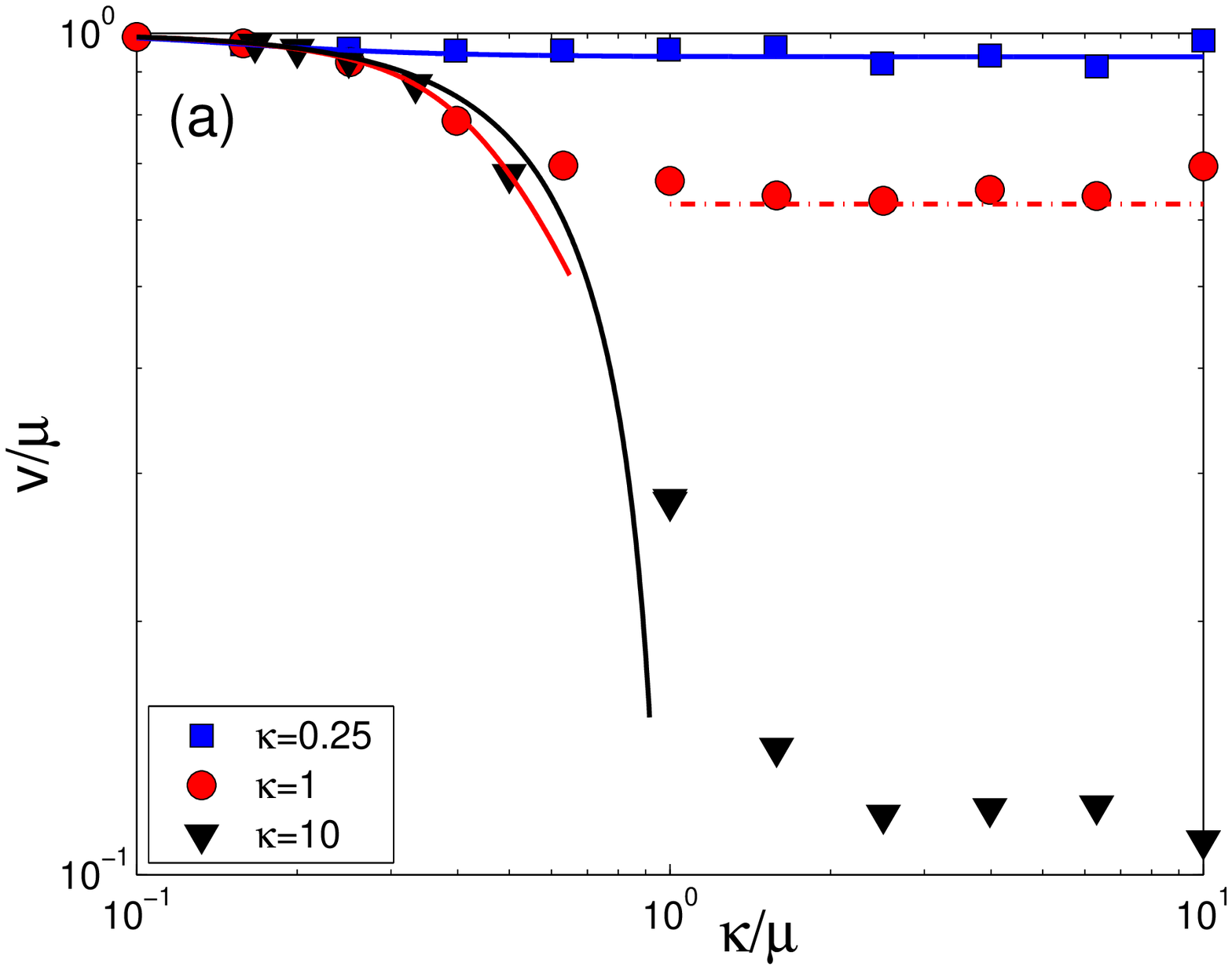,width=8.0cm}
 \epsfig{figure=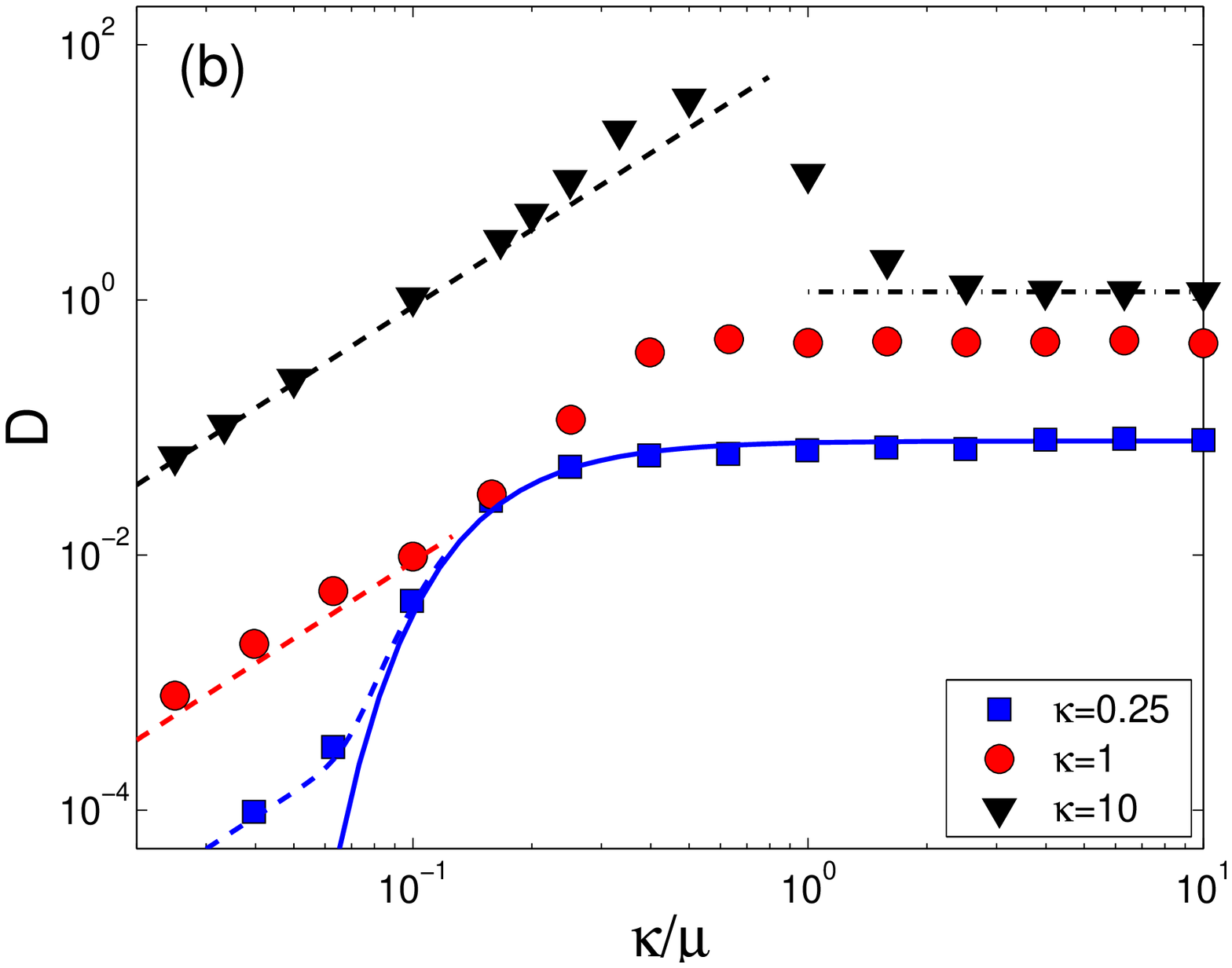,width=8.0cm}
\caption{Normalized mean velocity $v/\mu$ (a) and diffusion
coefficient $D$ (b) as functions of $\kappa/\mu$ for fixed values of
$\kappa$. In panel (a), solid lines show equation (\ref{meanvel})
for the mean velocity in the running regime, while the dot-dashed
line is the trapping regime model result (\ref{trapvel}). In panel
(b), the solid line is the leading-order perturbation result
(\ref{D1}) for $D$ in the running regime, while the dashed lines
include the order-$\kappa^4$ effects given in equation
(\ref{DFDC2}). The dot-dashed line is the trapping regime value of
$D$, given by equation (\ref{Dtrap}).
 }\label{fig1}
\end{figure}
Gaussian random functions may be constructed using a
combination of a large number $N$ of Fourier modes, as follows:
\begin{equation}
f(t) = \frac{1}{\sqrt{N}} \sum_{n=1}^N a_n \cos(\omega_n t)+b_n
\sin(\omega_n t). \label{numf}
\end{equation}
The amplitudes $a_n$ and $b_n$ are random numbers from a Gaussian
distribution of mean zero and unit variance. The random frequencies
$\omega_n$ are from a distribution shaped as the Fourier transform
of $R(t)$ --- for the correlation function (\ref{e7})  this means
the $\omega_n$ are also Gaussian distributed, with mean zero and
(nondimensional) variance one. The function $f(t)$ constructed in
this way is Gaussian in the limit $N\to \infty$
\cite{Kramer,Kraichnan}; we use $N=100$ in the numerical simulations
reported here.

In each realization, the independent noise functions $f(t)$ and
$g(t)$ are generated as above. The ordinary differential equation
(\ref{e8}) is then solved numerically, with initial condition
$x(0)=0$, to determine $x(t)$. By averaging over a large ensemble
(typically $10^4$) of realizations, we determine the time-dependent
moments $\left<x(t)\right>$ and $\left<x^2(t)\right>$, and  the
variance $\text{var}(x)$. A linear fit to the average position and
variance at large $t$ (we use nondimensional end-times of $t=10$ to
$t=50$ for this fitting) gives, respectively, the mean velocity $v$
and the diffusion constant $D$ from the slope of the fitted lines,
in accord with the definitions (\ref{vasym}) and (\ref{Ddefn}).
\begin{figure}
\centering
 \epsfig{figure=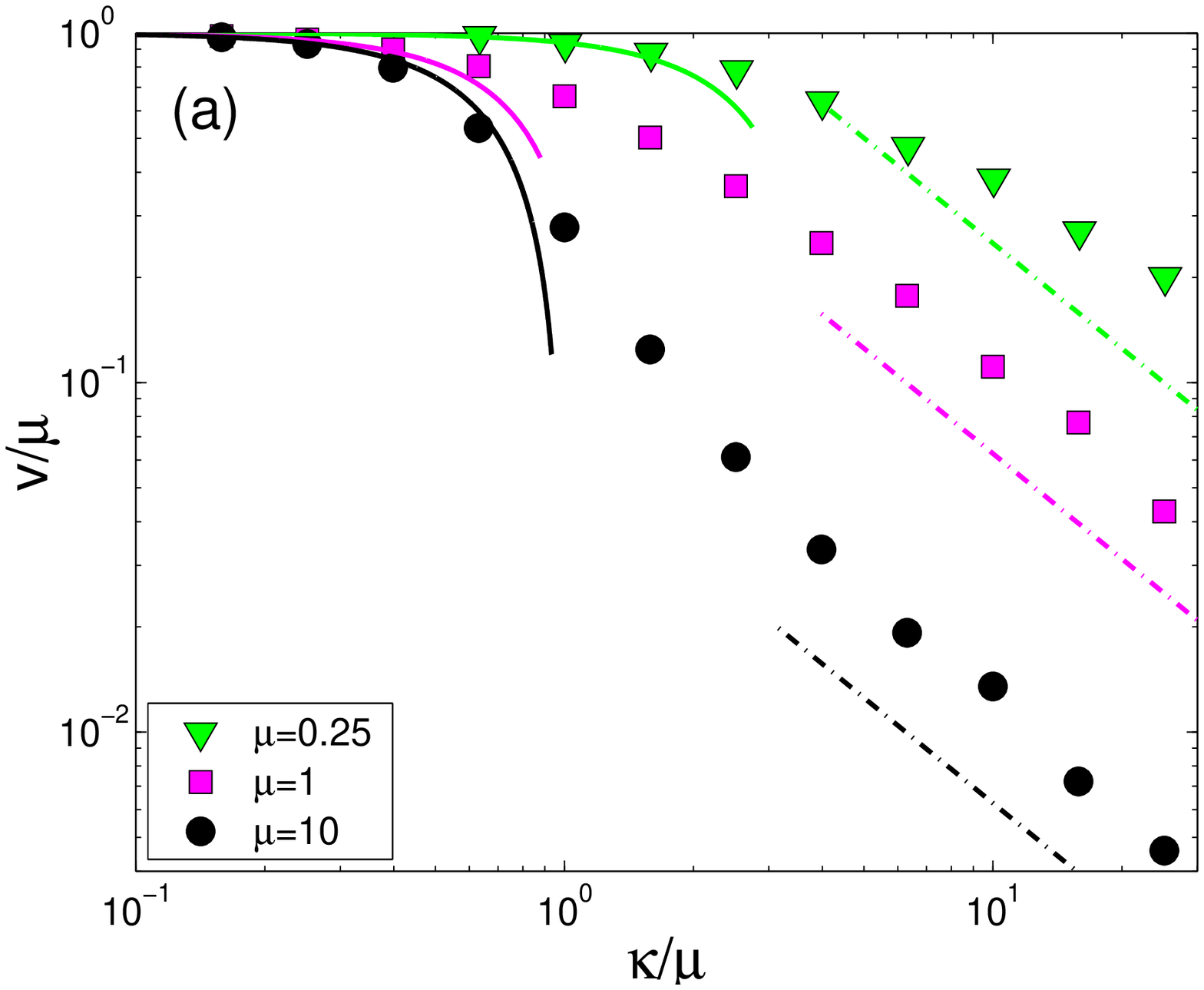,width=8.0cm}
 \epsfig{figure=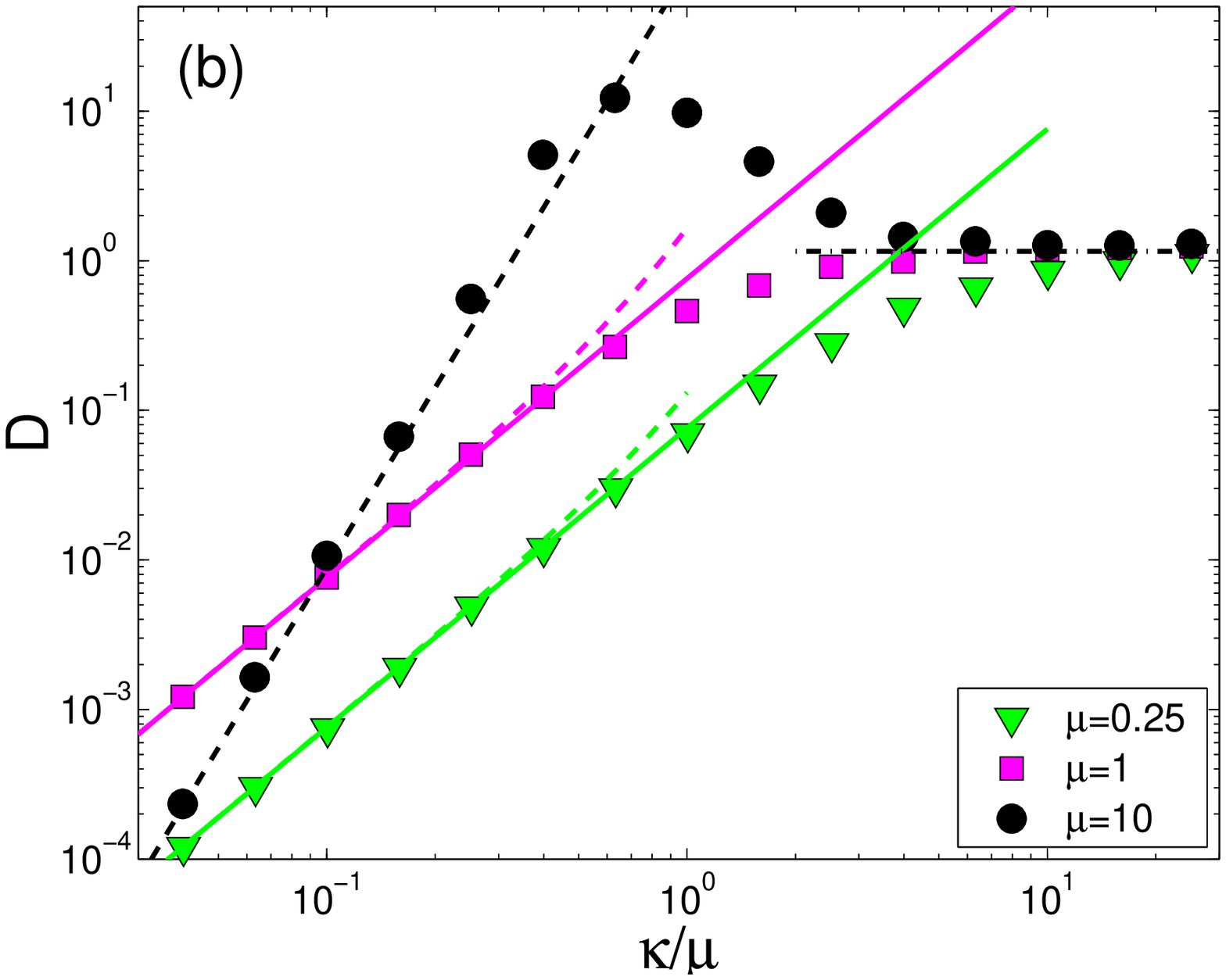,width=8.0cm}
\caption{Normalized mean velocity $v/\mu$ (a) and diffusion
coefficient $D$ (b) as functions of $\kappa/\mu$ for fixed values of
$\mu$. In panel (a), solid lines show equation (\ref{meanvel}) for
the mean velocity in the running regime, while the dot-dashed lines
are the trapping regime model result (\ref{trapvel}). In panel (b),
the solid lines are the leading-order perturbation  results
(\ref{D1}) for $D$ in the running regime, while the dashed lines
include the order-$\kappa^4$ effects given in equation
(\ref{DFDC2}). The dot-dashed line is the trapping regime value of
$D$, given by equation (\ref{Dtrap}).}\label{fig1b}
\end{figure}

Results from these numerical simulation are plotted as symbols in
Figures \ref{fig1} and \ref{fig1b}. Note the same symbol types are
 used as in the schematic parameter plane of Figure
\ref{figparamplane}, with the constant-$\kappa$ results shown in
Figure \ref{fig1}, and the constant-$\mu$ results in Figure
\ref{fig1b}. We choose to plot all results as functions of
$\kappa/\mu$ in order to highlight several important scalings. Panel
(a) of each figure shows the normalized mean velocity $v/\mu$, while
the diffusion coefficient $D$ appears in the  panel (b). Note all
scales are logarithmic. It appears that two distinct regimes of
simple scalings arise: broadly speaking these regimes appear to
correspond to $\kappa/\mu$ being respectively much smaller than, and
much larger than, unity. However, as we show in sections \ref{sect5}
and \ref{sect6} below the detailed picture is somewhat more
complicated, with the border between the running regime (section
\ref{sect5}) and the trapping regime (section \ref{sect6}) having
the shape shown in the schematic Figure \ref{figparamplane}.
Analytical results are derived below for limiting behaviour in each
regime and these are shown as the various curves in Figures
\ref{fig1} and \ref{fig1b}.

\section{Running regime and perturbation method}
\label{sect5}

When the mean velocity $\mu$ is sufficiently large, or the noise
parameter $\kappa$ is sufficiently small, particles moving in the
fluctuating potential according to equation (\ref{e8}) may be
treated as slightly perturbed from their deterministic paths. In
this parameter regime (to be defined in detail below) particles
typically move with a speed close to $\mu$ and move swiftly over the
potential landscape (i.e., covering many spatial periods during the
fluctuation timescale of the potential); we therefore call this the
\emph{running regime} of parameters. This regime is of particular
interest to designers of oscillator circuits and some of the results
of this section have been previously derived (using different
methods) in \cite{ODoherty_IEEE07}.

To study this regime more carefully, it is helpful to move with a
reference frame traveling at the deterministic mean velocity $\mu$.
By introducing the new variable $y(t)$ defined as
\begin{equation}
y(t)=x(t)- \mu \,t,
\end{equation}
equation (\ref{e8}) may be rewritten as
\begin{equation}
\dot y = \kappa \left[f(t) \cos\left( y+\mu t\right) +
 g(t) \sin\left( y+\mu t\right)   \right]. \label{ye1}
 \end{equation}
We combine the noise sources and time-oscillatory terms to create
two new stationary Gaussian random processes $F(t)$ and $G(t)$,
defined by
\begin{eqnarray}
F(t) &=& f(t) \cos \mu t + g(t) \sin \mu t, \nonumber\\
G(t) &=& -f(t) \sin \mu t + g(t) \cos \mu t.
\end{eqnarray}
From these definitions, it is straightforward to verify that $F(t)$
and $G(t)$ each have mean zero, and correlation function
\begin{equation}
\left<F(t)F(0)\right> = \left<G(t)G(0)\right> = R(t) \cos \mu t.
\label{Fe1}
\end{equation}
However, unlike $f(t)$ and $g(t)$, the new source terms $F(t)$ and
$G(t)$ are not independent, since their cross-correlation is given
by
\begin{equation}
\left< F(t) G(0) \right> = -\left< G(t) F(0) \right> = R(t) \sin \mu
t.
\end{equation}
In terms of the new noise processes, the equation of motion
(\ref{ye1}) is
\begin{equation}
 \dot y = \kappa \left[F(t) \cos y + G(t) \sin y \right].
\label{ye2}
 \end{equation}

Next, we assume a regular perturbation expansion for $y(t)$ in
powers of the parameter $\kappa$. The results of such an approach
will be useful if the distance traveled by the particle in the
characteristic time for the potential to decorrelate is
significantly less than the spatial period of the potential.
Provided this condition holds, the errors in approximating $\cos y$
and $\sin y$ by their series expansions about $y=0$ do not
accumulate over time. In terms of the dimensionless parameters
introduced in section \ref{sect3}, the spatial period of the
potential is of order one. The decorrelation time of the potential
felt by the particle in the moving reference frame depends on $\mu$;
from (\ref{Fe1}) we can estimate this as $\text{min}(1,1/\mu)$. The
distance traveled by the particle during this time is of order
$\kappa \,\text{min}(1,1/\mu)$ and the condition for the
perturbation expansion to hold is therefore expressible as $\kappa
\,\text{min}(1,1/\mu) \ll 1$, or equivalently as
\begin{equation}
\kappa \ll \text{max}(1,\mu). \label{runcond}
\end{equation}
We adopt this condition as defining the running regime of parameter
values.

 We proceed to assume a regular perturbation expansion for
$y(t)$ for the $\kappa\to 0$ limit of the form
\begin{equation}
y(t) = \kappa y_1(t) + \kappa^2 y_2(t)+\kappa^3 y_3(t)+\ldots,
\end{equation}
where each $y_i(t)$ is a stochastic process. Substituting this into
equation (\ref{ye2}), expanding the trigonometric terms, and
gathering powers of $\kappa$ yields the following system of
equations for the unknown functions $y_1(t)$, $y_2(t)$, and
$y_3(t)$:
\begin{eqnarray}
\dot y_1 &=& F(t) \label{y1eqn}\\
\dot y_2  &=& G(t) y_1 \label{y2eqn}\\
\dot y_3 &=& -\frac{F(t)}{2} y_1^2 + G(t) y_3.\label{y3eqn}
\end{eqnarray}
\subsection{Mean velocity in running regime}
The solution of (\ref{y1eqn}) with initial condition $y(0)=0$ is
\begin{equation}
y_1(t) = \int_0^t F(t_1) dt_1 \label{e36}
\end{equation}
and substitution into (\ref{y2eqn}) yields
\begin{equation}
\frac{d y_2(t)}{d t} = \int_0^t G(t) F(t_1) dt_1. \label{e37}
\end{equation}
Ensemble-averaging equations (\ref{e36}) and (\ref{e37}) leads to
\begin{eqnarray}
\left< y_1(t)\right> &=& 0 , \nonumber\\
\frac{d}{d t}\left<  y_2(t)\right> &=& - \int_0^t
R(t-t_1)\sin\mu(t-t_1) d t_1 \nonumber\\
&=& -\int_0^t R(s) \sin(\mu s) \,d s,
\end{eqnarray}
where the substitution $s=t-t_1$ has been used. This result implies
that in the long-time limit $t\to \infty$, we obtain a finite mean
value for  $d\left< y\right>/dt$ at leading order $\kappa^2$, given
by
\begin{equation}
\lim_{t\to\infty}\frac{d }{d t}\left< y\right> = -\kappa^2
\int_0^\infty R(s)\sin(\mu s)\, ds + O(\kappa^4).
\end{equation}
The mean velocity defined in equation (\ref{vdefn}) is therefore
given to leading order (as $\kappa\to 0$) by
\begin{equation}
v=\lim_{t\to\infty}\frac{d }{d t}\left< x\right> =\mu -\kappa^2
\int_0^\infty R(s)\sin(\mu s)\, ds + O(\kappa^4). \label{meanvel}
\end{equation}
The normalized mean velocity $v/\mu$ given by this result is plotted
as a solid line in panels (a) of Figures \ref{fig1} and \ref{fig1b}.
It matches numerical results well within the running regime defined
by equation (\ref{runcond}).

\subsubsection{Large-$\mu$ asymptotics of mean velocity}
To investigate the asymptotic behavior of (\ref{meanvel}) at large
$\mu$  we integrate by parts:
\begin{equation}
\int_0^\infty R(s)\sin(\mu s)\, ds = -\left[R(t)
\frac{1}{\mu}\cos(\mu t)\right]_{t=0}^{t\to \infty} + \frac{1}{\mu}
\int_0^\infty R^\prime(s) \cos (\mu s)\, ds.
\end{equation}
For sufficiently smooth correlation functions $R(t)$, this
integration by parts may be repeated in an iterative fashion, to
obtain an asymptotic series is inverse powers of $\mu$
\cite{BenderOrszag}. To leading order in $1/\mu$, we obtain
\begin{equation}
\int_0^\infty R(s) \sin(\mu s) \, ds \sim \frac{1}{\mu} +
O\left(\frac{1}{\mu^2}\right) \quad\text{ as }\mu\to \infty
\end{equation}
and we conclude that deviations of the mean velocity from $\mu$ are
of order $\kappa^2/\mu$ in this regime of parameter values.
\subsection{Diffusion coefficient in running regime}
The variance of $y(t)$ (which equals the variance of $x(t)$) is
given to leading order by
\begin{equation}
\text{var }y = \kappa^2 \left< y_1^2(t)\right>+O(\kappa^4),
\end{equation}
and from (\ref{e36}) we obtain
\begin{equation}
\left< y_1^2(t)\right> =  \int_0^t dt_1 \int_0^t dt_2 \, R(t_1-t_2)
\cos \mu(t_1-t_2).
\end{equation}
The slope of $\text{var }y(t)$ is given by its derivative with
respect to $t$:
\begin{equation}
\frac{d}{d t} \text{var }y = 2 \kappa^2\int_0^t R(s) \cos (\mu s)\,
ds + O(\kappa^4)
\end{equation}
and as $t\to\infty$ this reaches the finite limit giving $D$ from
equation (\ref{Dlim}):
\begin{equation}
D = \kappa^2\int_0^\infty R(s) \cos(\mu s)\, ds +O(\kappa^4).
\label{D1}
\end{equation}
This leading-order result for $D$ is plotted with solid lines in
panel (b) of Figures \ref{fig1} and \ref{fig1b}. While it matches
numerical results well for low $\kappa$ and low $\mu$ cases, it is
noticeable that even well within the running regime (with $\kappa
\ll \mu$) it is very inaccurate (not even appearing on the scale of
the Figures) for $\kappa=10$ (Fig.~\ref{fig1}(b)) and for $\mu=10$
(Fig.~\ref{fig1b}(b)). The reason for this loss of accuracy is
examined in detail in the following section.
\subsubsection{Large-$\mu$ asymptotics of diffusion coefficient}
We consider the asymptotic behavior of the diffusion constant at
large values of $\mu$. For correlation functions $R(t)$ which are
sufficiently smooth at $t=0$, the order $\kappa^2$ integral in
equation (\ref{D1}) can decrease rapidly with increasing $\mu$. For
example,  using the  correlation function of equation (\ref{e7}) in
equation (\ref{D1}) leads to the prediction that the diffusion
constant is exponentially small at large $\mu$ values:
\begin{equation}
D =\kappa^2\sqrt{\frac{\pi}{2}} e^{-\frac{\mu^2}{2}} + O(\kappa^4)
\text{  as  } \mu\to \infty. \label{Dexpsmall}
\end{equation}
Because the first non-vanishing term in the perturbation expansion
for $D$ may be exponentially small, we proceed to calculate to
higher orders. The contribution to $\text{var }y$ at order
$\kappa^4$ is given by
\begin{equation}
\left< y_2^2\right> + 2\left< y_1 y_3\right>- \left<y_2\right>^2,
\end{equation}
and after extensive algebraic manipulation this can be shown to
yield a contribution to the diffusion constant $D$ of
\begin{eqnarray}
D_2&=&\kappa^4 \int_0^\infty \!ds_1\!\int_0^\infty
\!ds_2\!\int_0^\infty\! ds_3 \nonumber\\
&&\left\{ R(s_1+s_2+s_3)R(s_2)\left[ -3
\cos \mu(s_1+s_2+s_3)\cos \mu s_2 + \sin\mu(s_1+s_2+s_3)\sin \mu s_2 \right]+\right.\\
&&\left. R(s_1+s_2)R(s_2+s_3)\left[ -2 \cos \mu(s_1+s_2)\cos \mu
(s_2+s_3) + 2\sin\mu(s_1+s_2)\sin \mu (s_2+s_3) \right]\right\}.
\label{D2}
\end{eqnarray}
The large-$\mu$ asymptotic behavior of this term may be calculated
(in a similar fashion to the mean velocity case above) as
\begin{equation}
D_2\sim \frac{\kappa^4}{\mu^2}\int_0^\infty R^2(t)dt \quad\text{ as
}\quad \mu\to\infty. \label{D2b}
\end{equation}
For the correlation function of equation (\ref{e7}) we therefore
have the following small-$\kappa$, large-$\mu$ asymptotic behavior:
\begin{equation}
D=
\kappa^2\sqrt{\frac{\pi}{2}}e^{-\frac{\mu^2}{2}}+\frac{\kappa^4}{\mu^2}
\frac{\sqrt{\pi}}{2}+O\left(\kappa^6,\mu^{-4}\right) \quad\text{ as
}\quad \mu\to\infty, \,\kappa\to 0. \label{DFDC2}
\end{equation}
This corrected form for $D$ is plotted with dashed lines in
Fig.~\ref{fig1}(b) and Fig.~\ref{fig1b}(b) and shows excellent
agreement with numerics within the running regime. We conclude that
the deviation from the leading-order result (\ref{D1}) noted above
is due to the order-$\kappa^4$ effects dominating the
order-$\kappa^2$ term when $\mu$ is sufficiently large.
\subsection{Summary of running regime}
In brief,  condition (\ref{runcond}) defines the running regime of
parameter values  and within this regime the mean velocity and
diffusion coefficient may be determined using a stochastic
perturbation expansion. To order $\kappa^2$ the mean velocity $v$ is
given in equation (\ref{meanvel}) and the diffusion coefficient in
equation (\ref{D1}). Asymptotic analysis for large $\mu$ reveals
that the velocity scales as $\mu-\kappa^2/\mu$ as $\mu\to\infty$,
while $D$ can have an order $\kappa^2$ term which is exponentially
small, see equation (\ref{Dexpsmall}). In this case the
order-$\kappa^4$ contribution of equation (\ref{D2}) will dominate.

\section{Trapping regime} \label{sect6}
In  section \ref{sect5} we analyzed particle motion under the
condition that $\kappa \ll \text{max}(1,\mu)$, which corresponds to
the running regime. At the other extreme, i.e. when $\kappa\gg
\text{max}(1,\mu)$, the dynamics of particles may be understood in
terms of trapping at low-velocity positions, with intermittent
releases and flights between trapping points. To analyze the
dynamics in this \emph{trapping regime}, it is useful to rewrite
equation (\ref{e8}) as
\begin{equation}
\frac{d x}{d t} = \mu + \kappa r(t) \sin\left( x+\phi(t)\right).
\label{trap1}
\end{equation}
The  \emph{phase} $\phi(t)$ and the \emph{amplitude} $r(t)$ are each
random functions of time, and are given in terms of the original
noise processes $f(t)$ and $g(t)$ as
\begin{equation}
\phi(t) = \tan^{-1} \frac{f(t)}{g(t)}, \quad \quad
r(t)=\sqrt{f(t)^2+g(t)^2}; \label{rphi}
\end{equation}
conversely, the noise sources may be expressed as $f(t)=r \sin \phi$
and $g=r \cos \phi$.

\subsection{Diffusion coefficient in trapping regime}
 Taking the limits $\mu\to 0$ and $\kappa \to \infty$ in
equation (\ref{trap1}) allows us to derive some useful results. In
this limit, the particle trajectories $x(t)$ are locked to the phase
variable: $x(t)=-\phi(t)$ (up to the addition of multiples of
$2\pi$), and in particular the velocity of the particle is given
explicitly as a function of time by $\dot x(t) = -\dot \phi(t)$. The
phase velocity $\dot \phi (t)$ in systems of this type has been
extensively studied \cite{Gleeson_PRE05, Gleeson_EPL06}, and known
results may be applied to find the diffusion coefficient in this
limit \cite{Gleeson_PhysicaA05}. The mean value of  $\dot \phi$ is
zero, and the phase velocity correlation function may be expressed
in terms of the correlation function $R(t)$ of the noise sources as
\begin{equation}
L(t)=\left<\dot{\phi}(0) \dot{\phi}(t)\right> = \frac{1}{2
R(t)^2}\left[R(t)R^{\prime\prime}(t)-R^\prime(t)^2 \right]
\log\left[ 1 - R(t)^2\right]. \label{Leqn}
\end{equation}
In this trapping limit the effective diffusion coefficient for the
particles may therefore be approximated by setting $x(t)=-\phi(t)$
and proceeding as follows:
\begin{eqnarray}
D =\frac{1}{2}\lim_{t\to\infty} \frac{d}{d t}\left< x^2(t)\right>
&=& \lim_{t\to\infty}\left<
\phi(t)\dot{\phi}(t)\right> \nonumber\\
&=& \lim_{t\to\infty} \left<\int_0^t \dot{\phi}(t_1) \dot{\phi}(t)\,
dt_1 \right>
\nonumber\\
&=&  \int_0^\infty L(t_1)\, dt_1 \label{Dtrap}
\end{eqnarray}
Given the noise correlation function, this integral may be
calculated numerically after applying equation (\ref{Leqn});  for
the case $R(t)=\exp(-t^2/2)$ this yields the value $D=1.158$. This
value is plotted with a dot-dashed line in Figures \ref{fig1}(b) and
\ref{fig1b}(b) and agrees with the numerical results for parameters
well within the trapping regime $\kappa \gg \text{max}(1,\mu)$.

\subsection{Mean velocity in trapping regime}\label{toysect}
In the strong trapping limit of $\mu\to 0$ the mean velocity of the
particles is necessarily zero. Numerical results (e.g.
Fig.~\ref{fig1b}(a)) indicate that if $\mu$ is non-zero but small
then the mean velocity scales linearly with $\mu/\kappa$, and we are
motivated to study the behavior of equation (\ref{trap1}) as a
small-$\mu$ perturbation about the strong-trapping case considered
above. As we shall see, it is a challenging problem to describe the
dynamics in this case, because of short (in time) but large (in
space) intermittent escapes of the particle from the strong-trapping
solution. As a consequence, we will revert to a simpler toy problem
in an attempt to understand the observed numerical scaling of the
mean velocity in this regime. We begin by giving the exact equation
for the deviation $z(t)=x(t)+\phi(t)$ of the particle position
$x(t)$ from its strong-trapping limit $-\phi(t)$:
\begin{equation}
\frac{d z}{d t}= \mu + \dot \phi(t)+ \kappa\, r(t) \sin z .
\end{equation}
Analysis of this equation is significantly complicated by the
dynamics of the non-Gaussian phase velocity $\dot \phi(t)$. For
example, noting that the phase velocity may be expressed as
\begin{eqnarray}
\dot \phi &=& \frac{d}{dt} \tan^{-1}\frac{f}{g} \nonumber\\
&=&\frac {g \dot f - f \dot g}{f^2+g^2}\nonumber\\
&=& \frac{\dot f \cos\phi - \dot g \sin\phi }{r},
\end{eqnarray}
and that $\dot f(t)$ and $\dot g(t)$ are independent Gaussian
processes, we see that the one-time distribution of $\phi(t)$ is the
same as that of $\s(t)/r(t)$, where $\s(t)$ is a zero-mean and unit
variance Gaussian random function and $r(t)$ is the amplitude given
by equation (\ref{rphi}) above. This fact may be used to find the
distribution of $\phi(t)$ in closed form \cite{Gleeson_PhysicaA05},
but here we use it to develop an analytically-tractable toy model
which demonstrates similar behavior to the full dynamics.

We consider splitting time into intervals of length $T$, and in each
interval replacing the time-varying functions $r(t)$ and $\dot
\phi(t)$ with random constants chosen from the appropriate one-time
distributions. This \emph{frozen-noise} approximation thus neglects
all variation of the phase and amplitude variables on timescales
shorter than $T$, but it allows the dynamics of the corresponding
$z(t)$ to be found be solving the constant-coefficient equation
\begin{equation}
\frac{d z}{d t}= \mu + \frac{\s}{r}+ \kappa r \sin z \label{toye}
\end{equation}
within each $T$-interval. Here $r$ is given by $r=\sqrt{f^2+g^2}$
and $\dot \phi$ is approximated by $\s/r$, with $f$, $g$, and $\s$
being independent Gaussian random numbers with mean zero and unit
variance, chosen anew for each new time interval.

We focus on the mean velocity $\left< \dot x(t)\right>$, which is
equal to $\left<\dot z(t)\right>$ since $\dot \phi$ has zero mean.
The ensemble-averaged $z$-velocity within the toy model is
calculated by solving (\ref{toye}) within each $T$-interval, and
then averaging over the possible values of $r$ and $\s$. There are
essentially two cases to consider.  If the random numbers $\s$ and
$r$ satisfy the inequality
\begin{equation}
\left(\mu+\frac{\s}{r}\right)^2-\kappa^2 r^2< 0\label{ineq}
\end{equation}
then a steady state solution of (\ref{toye}) exists for $z(t)$, and
so  after a transient  the particle reaches this steady state and
the velocity decays to zero. For sufficiently long time intervals
$T$, we can neglect the transient and so consider the effective
velocity to be zero in cases satisfying (\ref{ineq}). If the
inequality is not satisfied, however, then no steady state solution
of equation (\ref{toye}) exists, and these \emph{running} solutions
have a time-averaged velocity calculated (see the appendix) to be
\begin{equation}
v_\text{run}=\pm\sqrt{\left(\mu+\frac{\s}{r}\right)^2-\kappa^2 r^2}.
\label{vrun}
\end{equation}
To determine the ensemble mean velocity it remains only to average
these running velocities over the distribution of $r$ and $\s$
values satisfying the inequality (\ref{ineq}). After some
manipulation (see appendix) it is possible to show that the
resulting mean velocity is linear in $\mu$, taking the form
\begin{equation}
\left< \dot z \right> = \frac{1}{2}\sqrt{\frac{\pi}{2}}
\frac{\mu}{\kappa} + O(\mu^2)\quad\text{  as }\frac{\mu}{\kappa}\to
0. \label{trapvel}
\end{equation}
Using this form for the mean velocity gives (after normalizing by
$\mu$) the dot-dashed curves in Figures \ref{fig1}(a) and
\ref{fig1b}(a). It is apparent from Fig.~\ref{fig1b}(a) that the
linear dependence on $\mu/\kappa$ in (\ref{trapvel}) qualitatively
matches the numerical results, but the prefactor is too small by a
factor of approximately 2. Given the approximations made in the
simplification to the toy problem used here, this level of agreement
is considered rather satisfactory.

\section{Discussion} \label{sect7}
To summarize: we have examined the behavior of first and second
moments of the solution $x(t)$ of the equation (\ref{en})
(nondimensionalized to equation (\ref{e8})) with independent random
forcing terms $f(t)$ and $g(t)$. Numerical simulation results are
supported by asymptotic analysis in the two main regimes of the
$\kappa$--$\mu$ parameter plane, termed the running regime (section
\ref{sect5}) and the trapping regime (section \ref{sect6})---see
also Figure \ref{figparamplane} and definition (\ref{runcond}).

The smoothness of the noise correlation function $R(t)$ at $t=0$ was
assumed at the beginning of section \ref{sect3}. If the smoothness
property $R^\prime(0)=0$ is violated by, for example, using
Ornstein-Uhlenbeck processes for $f(t)$ and $g(t)$, then the
analysis of the trapping regime which relies on $\dot f(t)$ and
$\dot g(t)$ being finite-variance processes is not valid. The
perturbation analysis in the running regime is unchanged, but note
the leading-order diffusion term (\ref{D1}) typically scales as
$\kappa^2/\mu^2$ for small $\kappa$ and large $\mu$ if $R$ is
non-smooth and so the order-$\kappa^4$ term (\ref{D2b}) will not
dominate as it does when $R$ is smooth.

It is interesting to compare the results for our model with related
work on noisy transport in periodic potentials. The non-monotone
dependence of the diffusion coefficient on $\kappa/\mu$, seen in
Fig.~\ref{fig1}(b) for $\kappa=10$ and in Fig.~\ref{fig1b}(b) for
$\mu=10$, is reminiscent of Figure 1 of \cite{Reimann_PRL01}. The
authors of  \cite{Reimann_PRL01} consider overdamped motion in a
tilted, spatially-periodic potential which (unlike our
randomly-fluctuating case) is static in time, but to which white
noise effects (not included by us) are added. They find the
diffusion coefficient in their model is greatly increased at a
critical value of the tilt parameter corresponding to the onset of
deterministically running solutions. We find the peak value of $D$
in our model occurs for $\kappa/\mu$ ratios of order unity (actually
approximately 0.5 to 0.7), which corresponds to the border between
the running and trapping regimes at high values of $\kappa$ or
$\mu$. Because particles may be either running or trapped in any
given realization, the overall variance of $x(t)$ across the
ensemble is particularly large for $\kappa/\mu \approx 1$, leading
to a diffusion coefficient which exceeds that found in cases where
each single realization is more typical of the ensemble average.

Our perturbation expansion method for the running regime bears some
resemblance to that used in \cite{Jansons_Lythe} to examine Stokes
drift in a fluid due to the presence of periodic waves. Substitution
of the Fourier expansions (\ref{numf}) for $f(t)$ and $g(t)$ in
equation (\ref{e8}) allows the equation of motion to be written as
\begin{equation}
\frac{d x}{d t} = \mu +  \frac{\kappa}{\sqrt{N}} \sum_{n=1}^N A_n
\cos (x-\omega_n t)+ B_n \sin(x-\omega_n t),
\end{equation}
with the sum admitting the interpretation as a superposition of
waves of various amplitudes (and directions) and random frequencies
$\omega_n$. Our result (\ref{meanvel}) may therefore be compared (by
taking $\mu=0$) to the drift velocity determined by a perturbation
exapansion in \cite{Jansons_Lythe}. Unlike  \cite{Jansons_Lythe},
however, the randomness of our waves means there is no preferred
wave direction, and so we find a zero drift velocity (we also do not
consider the additive white noise diffusion effects used in
\cite{Jansons_Lythe} and \cite{VandenBroeck}). The directional
symmetry of the random wave field is broken when $\mu$ is non-zero
and in this case the deviation of $v$ from the deterministic
velocity $\mu$ is of order $\kappa^2$ (from equation
(\ref{meanvel})), which (like in \cite{Jansons_Lythe}) is the square
of the wave amplitude. It would be of considerable interest to
extend our methods to include the effects on the particle
trajectories of additive white noise terms as in
\cite{Reimann_PRL01}, \cite{Jansons_Lythe}, \cite{VandenBroeck}, in
addition to the multiplicative colored noise terms considered here.

Finally, we note the possibility of using spatially-periodic
potentials for sorting of microparticles, which has attracted
considerable recent attention \cite{Korda}, \cite{MacDonald},
\cite{Gop}, \cite{Gleeson_sorting}. If experimental periodic
potentials can also be made to fluctuate randomly in time (perhaps
by modulation of the laser intensities generating optical lattices
as in \cite{Korda}, \cite{MacDonald}) then the results of this paper
would provide a useful basis for predicting mean velocities and
spreads of particles being sorted within the device.
\section*{Acknowledgements}
 This work is funded by Science Foundation Ireland under
Investigator Award 06/IN.1/I366  and Research Frontiers Programme
05/RFP/MAT0016. The author benefited from helpful discussions with
Fergal O'Doherty and Grant Lythe.

\appendix
\section*{Appendix: Mean velocity in toy model for trapping regime}
In section \ref{toysect} we introduced a simplified toy model for
the trapping regime within which we can gain some analytical insight
into the behavior of the mean velocity. Recall equation (\ref{toye})
generates a non-zero contribution to the mean velocity $\left< \dot
z \right>$ in a given $T$-interval if the parameters
$r=\sqrt{f^2+g^2}$ and $\sigma$ do not satisfy inequality
(\ref{ineq})---here $f$, $g$ and $\sigma$ are unit variance, mean
zero normal random numbers in each $T$-interval. Indeed, in this
case equation (\ref{toye}) immediately yields an integral solution
for the time taken to travel the spatial period $2\pi$:
\begin{equation}
\int_0^{2\pi} \frac{dz}{\mu + \frac{\s}{\kappa}+\kappa r \sin z} =
\int_0^{2\pi/v_{\text{run}}} dt,
\end{equation}
which gives the equation (\ref{vrun}) in the main text for the
running velocity $v_\text{run}$ in that particular $T$-interval
(assuming $T$ is sufficiently large). In all cases the sign of
$v_\text{run}$ is the same as the sign of the quantity
$\mu+\frac{\s}{r}$.

Next, we consider the averaging of (\ref{vrun}) over the possible
values of $r$ and $\s$. The values of $\s$ are chosen from the range
$(-\infty,\infty)$ with density $e^{-\s^2/2}/\sqrt{2\pi}$, while the
values of $r$ are from the range $[0,\infty)$ with density $r
e^{-r^2/2}$. When averaging over all values of $r$ and $\s$
satisfying the inequality
\begin{equation}
\left(\mu+\frac{\s}{r}\right)^2-\kappa^2 r^2> 0\label{ineq2}
\end{equation}
we must consider three regions of the $r$--$\s$ parameter plane, as
follows:
\begin{itemize}
\item Region 1: $\s=0$ to $\infty$, with $r=0$ to
$(\mu+\sqrt{\mu^2+4\kappa \s})/2\kappa$. In this region
$v_\text{run}$ is positive.
\item Region 2: $\s=-\mu^2/4 \kappa$ to $0$, with $r=(\mu-\sqrt{\mu^2+4\kappa \s})/2\kappa$ to
$(\mu+\sqrt{\mu^2+4\kappa \s})/(2\kappa)$. In this region
$v_\text{run}$ is positive.
\item Region 3: $\s=-\infty$ to $0$, with $r=0$ to
$(-\mu+\sqrt{\mu^2-4\kappa \s})/2\kappa$. In this region
$v_\text{run}$ is negative.
\end{itemize}
Because we are interested only in the leading-order behavior of
$\left< \dot z \right>$ at small deterministic velocity $\mu$, we
focus on the linear term in an expansion in small $\mu$:
\begin{equation}
\left< \dot z \right> \approx\left. \mu \frac{\partial \left< \dot z
\right>}{\partial \mu}\right|_{\mu=0}+O\left(\mu^2\right).
\end{equation}
It is easy to show that Region 2 does not contribute to the mean
velocity at this order, while Regions 1 and 3 each contribute an
amount equal to
\begin{eqnarray}
&&\hspace{-0.6cm}\frac{\mu}{\sqrt{2\pi}}\int_0^\infty
\int_0^{\sqrt{\sigma/\kappa}} e^{-\frac{\s^2}{2}}r
e^{-\frac{r^2}{2}} \left.\frac{\partial
v_\text{run}}{\partial \mu}\right|_{\mu=0} dr\, d\s \nonumber\\
&=&\frac{\mu}{\sqrt{2\pi}}\int_0^\infty
\int_0^{\sqrt{\sigma/\kappa}} e^{-\frac{\s^2}{2}}r
e^{-\frac{r^2}{2}} \frac{\s}{\sqrt{\s^2-\kappa^2 r^4}} \,dr\, d\s
\end{eqnarray}
After the change of variable $w=\kappa r^2/\s$ we obtain
\begin{eqnarray}
&&\hspace{-0.6cm} \frac{\mu}{\sqrt{2\pi}} \frac{1}{2\kappa}
\int_0^\infty \int_0^1 \s e^{-\frac{\s^2}{2}} e^{-\frac{\s
w}{\kappa}} \frac{1}{\sqrt{1-w^2}}\,
dw\, d\s \nonumber\\
&\sim& \frac{\mu}{4 \kappa}\sqrt{\frac{\pi}{2}}+
o\left(\frac{1}{\kappa}\right) \quad\text{ as}\quad \kappa\to
\infty,
\end{eqnarray}
where the final expression has been obtained by taking $\kappa\to
\infty$ in the integrand. Adding the equal contributions from
Regions 1 and 3 then gives the asymptotic expression of equation
(\ref{trapvel}) of the text for the averaged velocity.

\end{document}